\documentclass[prd]{revtex4}
\usepackage{amsfonts}
\usepackage{amsmath}
\usepackage{amssymb}
\usepackage{bbm}
\usepackage[utf8]{inputenc}
\usepackage[caption=false]{subfig}
\usepackage{tensor}
\usepackage{slashed}
\usepackage[vcentermath]{youngtab}
\usepackage{hyperref}
\usepackage{natbib}
\usepackage{enumerate}
\usepackage{braket,mleftright}
\usepackage[mathscr]{euscript}
\usepackage{cleveref}

\renewcommand{\ket}[1]{\ensuremath{\left |#1 \right\rangle}}
\renewcommand{\bra}[1]{\ensuremath{\left\langle#1\right|}}

 \newcommand{\refeq}[1]{(\ref{#1})}
\newcommand{\lagrangian}{\mathcal{L}}

\newcommand{\so}{\ensuremath{\mathfrak{so}}}

\newcommand{\su}{\ensuremath{\mathfrak{su}}}
\newcommand{\Oo}{\operatorname{O}}
\newcommand{\U}{\operatorname{U}}
\newcommand{\SO}{\operatorname{SO}}
\newcommand{\SU}{\operatorname{SU}}

\newcommand{\id}{\mathbbm{1}}

\newcommand{\procarep}{\ensuremath{ \left(\frac{1}{2}, \frac{1}{2}\right)}}
\newcommand{\diracrep}{\ensuremath{ \left(\frac{1}{2},0 \right) \oplus \left(0, \frac{1}{2}\right)}}
\newcommand{\lrep}[1]{\ensuremath{\left(#1\right)}}

\newcommand{\sschir}{\left(\frac{3}{2},0 \right) \oplus \left(0, \frac{3}{2} \right)}
\newcommand{\dschir}{\left(1,\frac{1}{2} \right) \oplus \left(\frac{1}{2},1 \right)}
\newcommand{\rsr}{\left(1,\frac{1}{2} \right) \oplus \left(\frac{1}{2},1 \right) \oplus \left(\frac{1}{2}, 0\right) \oplus \left(0,\frac{1}{2}\right)}
\newcommand{\egr}{\left(1,\frac{1}{2} \right) \oplus \left(\frac{1}{2},1 \right) \oplus \left(\frac{3}{2}, 0\right) \oplus \left(0,\frac{3}{2}\right)}

\providecommand{\U}[1]{\protect\rule{.1in}{.1in}}
\begin{document}

\title{Wave equations for spin $\frac{3}{2}$ quantum fields.}
\author{J. Escamilla--Mu\~noz}
\author{Selim G\'omez-\'Avila}
\affiliation{Área Académica de Matemáticas y Física, Instituto de Ciencias Básicas e Ingeniería, Universidad Autónoma del Estado de Hidalgo, Edificio MF, Carretera Pachuca-Tulancingo Km. 4.5, Mineral de la Reforma, Hidalgo, México}

\begin{abstract}
{In this work, we review  formulations of wave equations for spin-3/2 fields constructed from different Lorentz group representations. We analyze the Joss-Weinberg single-spin chiral representation and the double-spin chiral representation,  focusing on the structure of their covariant operators. We explore the Duffin-Kemmer-Petiau (DKP) formalism and its algebraic properties, originally introduced for spin-0 and spin-1 particles, and here considered as a potential framework for spin $3/2$. As a result, we recover the well--known Rarita-Schwinger representation and we find a new possibility in the $\egr$ representation.}
\end{abstract}

\maketitle


\section{Introduction.}
\label{sec:intro}
The Standard Model of particle physics is an interacting Quantum Field Theory, or QFT \citep{Weinberg:1995mt},  describing all known particles and interactions with remarkable accuracy \citep{Navas2024}. To do so, it employs only massive spin zero scalar fields, spin $\frac{1}{2}$ massless Dirac spinors, and helicity $\pm 1$ massless vector fields \citep{MCCABE2007i}. Other types of fields have neither been experimentally discovered nor ruled out on grounds of mathematical consistency or physical necessity. 
Despite its undeniable success, the Standard Model faces several challenges. It has many free dimensionless parameters exhibiting large hierarchies between them, and it does not account for neutrino oscillations, the matter–antimatter asymmetry of the universe, and the nature of dark matter and dark energy \citep{Ellis2012, ABDALLA202249}.

Solving these questions requires us to go beyond the Standard Model (BSM) and perhaps even beyond quantum field theory. This requires not only the phenomenological exploration of QFTs such as Grand Unified Theories, but also a systematic exploration of the limits and consistent possibilities permitted by the principles of quantum field theory. One such exploration is the construction of novel QFTs for the description of matter fields, which can then become ingredients in the phenomenology of BSM theories.

Spin $\frac{3}{2}$ particles exist as hadronic excitations in the QCD spectrum \citep{Anderson1952}. In supergravity scenarios, a massive gravitino with spin $\frac{3}{2}$ also appears upon the SSB of supersymmetry\citep{ELLIS1984410}. Historically, the description of such spin $\frac{3}{2}$ particles has been carried out in the Rarita-Schwinger vector--spinor formalism \citep{Rarita:1941mf}. However, this description is known to be inconsistent when subject to general interactions. 
\citep{Velo:1969bt}.

In the present work, several possible frameworks for the description of spin  $\frac{3}{2}$ particles are examined, and two novel possibilities are identified. 

\section{Representation theory for Poincaré quantum fields.}
\label{sec:poirep}
We know since seminal work by \citet*{Wigner1939} that elementary systems correspond to unitary irreducible representations of the Poincaré group $IO(1,3) \simeq { \mathbf {R} ^{1,3}\rtimes \Oo(1,3)\,} $.  A general transformation in this group can be written as: 
\begin{equation}
    g(b, \Lambda) = T(b)\, \Lambda,
\end{equation}
where $b$ represents a spacetime translation, and $\Lambda$ is a proper Lorentz transformation $\Lambda\in \Oo(1,3)$. 

Any field that transforms irreducible under the action of the operators of the full Poincaré group, will correspond to the idea of an elementary system. In particular, this means that it belongs to one of the orbits of Wigner's classification: 

\begin{enumerate}
    \item $m^2 > 0$, with well--defined values for the Casimir operators $\mathcal C_2 = P^2$ and $\mathcal C_4 = W^2$ with $W^\sigma = \frac{1}{2}\tensor{\epsilon}{^\sigma ^\rho^\mu^\nu} P_\rho \tensor{M}{_\mu_\nu}$ and eigenvalues $w^2 =- m^2 s(s+1) .$
    \item $m=0$ with $P^0 > 0$, with well defined value for the helicity $\lambda$ in $W^\mu + \lambda P^\mu = 0$. 
    \item $m=0$ y $P^\mu = 0$, which uniquely corresponds to what we call the vacuum representation. 
    \item $m^2 < 0$, with well--defined values for the Casimir operators $\mathcal C_2 = P^2$ and $\mathcal C_4 = W^2$ with $W^\sigma = \epsilon$. (These are called \emph{tachyonic} representations). 
\end{enumerate}
In the following we will center our attention in representations in the first orbit, that is, fields associated with massive particles. 

The spacetime translation comprise an invariant subgroup of the Poincaré group; the existence of such a subgroup in turn means that the Poincaré group is neither simple nor semisimple.  Its irreducible representations, in consequence, are built from the representations of the Lorentz subgroup as induced representations \citep{Raczka1986-be}. For this reason, we now turn to the representation theory of the Lorentz group.  

There is a well-known small dimension isomorphism between the Lorentz algebra $\so(1,3)$ and the direct sum $\su(2) \oplus \su(2)$. At the group level this manifests as
\begin{equation}
    \SO(1,3) \simeq \left( \SU(2) \times \SU(2)\right)/\mathbb{Z}_2.
\end{equation}
This isomorphism allows us to exploit the representation theory of $\SU(2)$; representations of $\SO(1,3)$ can be labeled by two semi-integers $(a,b)$. The irreducible representations of $\Oo(1,3)$ have two forms: $\lrep{a,b}$ for $a=b$, or $(a,b) \oplus (b,a)$ for $a\neq b$. Therefore, our Poincaré matter fields with definite $m^2$ and definite $W^2$ (or $s$) will also need a label $(a,b)$\ to tell us which $\Oo(1,3)$ representation was used to construct it. Fields with the same values for $m^2$ and $s$ are equivalent as free fields, but not necessarily as interacting fields. (We will use the symbol $(a,b)_\chi$ to denote the direct sum $(a,b) \oplus (b,a)$ if $a \neq b$, and $(a,b)_\chi = (a,b)$ when $a=b$). 

We can use a given $(a,b)_\chi$ representation to induce a $(m^2, j)$ Poincaré field if 
\begin{equation}
    j \in \{ |a-b|, |a-b| +1, \dots a + b\}.
\end{equation}

A Poincaré field in the massive orbit $(m^2, j)$ propagates $2j+1$ degrees of freedom. An $(a,b)_\chi$ representation containing $j$ will in general consist of more degrees of freedom, which requires the presence of constraints. As we will discuss below, wave equations are a covariant record of the way in which a Poincaré field is induced by some Lorentz representation and the unwanted degrees of freedom are projected out. As per  \citet{Weinberg:1964cn}, ``a free-field equation is nothing but an invariant record of which components are superfluous''.

In general, if we are interested in spin $j$ fields, we would like to have it as the highest spin contained in the $(a,b)_\chi$ representation. With this condition in mind, we can find spin $j$ in the representations
\begin{equation}
    \left(j,0\right)_\chi, \left(j-\frac{1}{2},\frac{1}{2}\right)_\chi, \left(j-1,1\right)_\chi, \left(j-\frac{3}{2},\frac{3}{2}\right)_\chi \dots 
\end{equation}

Relativistic QFT requires the careful management of constraints, with fundamentally result from the use of Poincaré representations induced by Lorentz representations.  Theories with many constraints quickly become unwieldy. In particular, even if they have the correct counting of degrees of freedom as free theories, the introduction of interactions can produce the propagation of \emph{ghosts}, that is, solutions with negative kinetic energy and the wrong spin \citep{Boulware1972}. An early example of this is the massive Proca and Rarita--Schwinger equations, where minimal coupling with an electromagnetic field was shown to produce effects such as superluminal propagation of some degrees of freedom as well as the loss of the canonical commutation rules \citep{Velo:1969bt, Johnson:1960vt}.

For this reason, we only  consider representations $(a,b)_\chi$ with at most two spin sectors. This means that we only have three possibilities for the description of spin $j$:
\begin{enumerate}
    \item Non--chiral representations $\left(\frac{j}{2},\frac{j}{2}\right)$ for $j=0,1$ only. 
    \item Single spin chiral representations $(j,0)_\chi$. 
    \item Double spin chiral representations $\left(j-\frac{1}{2},\frac{1}{2}\right)_\chi$,
\end{enumerate}
and of course, direct sums of these. For $j=0,\frac{1}{2}, 1$ these include all of the fields used in the Standard Model. 

In this note, we will review existing wave equations for spin $(3/2)$, the Joos-Weinberg single--spin chiral representation $\sschir$, the double--spin chiral representation $\dschir$, and consider the Rarita--Schwinger 
\begin{equation}
    \rsr,
\end{equation} and the novel 
\begin{equation}
    \egr
\end{equation} 
representations as generalizations for half--integer spin of the Duffin-Kemmer-Petiau (DKP) construction \citep{1936Petiau, Duffin1938, 1939Kemmer}.

\section{Covariant bases and wave equations}

Covariant relativistic wave equations are hyperbolic partial differential equations in space and time, with coefficients that have definite transformation properties under the Lorentz group. For a given $(a,b)_\chi$ it may be possible to construct several unequivalent wave equations. 

In order to classify all covariant relativistic wave equations or CRWEs for a representation on a vector space $V$, we need to characterize the general lineal group $gl(V)$ {with respect to the action of the Lorentz group}. We will use the notation 
\begin{equation}
    \ket{a_3,b_3}_{(a,b)} \equiv \ket{a,a_3}\otimes\ket{b,b_3},
\end{equation}
to denote the states, and omit the $(a,b)$ label when the context makes it unnecesary. 

For the non--chiral spin zero representation $(0,0)$, the vector space is one--dimensional, and spanned by the state
\begin{equation}
    \ket{0} \equiv \ket{0,0}_{(0,0)}.
\end{equation}
Similarly, the operator space is spanned by the projection
\begin{equation}
    \id = \ket{0}\bra{0} \equiv \ket{0,0}_{(0,0)} \bra{0,0}_{(0,0)}.
\end{equation}
In terms of the dimension of the $\SO(1,3)$ representations we get the rather trivial looking equation
\begin{equation}
    1 \otimes 1 \simeq 1.
\end{equation}
The only kinetic term that we can write using this and the momentum $P^\mu$ is $P^2$; this is how we end up with Klein-Gordon's equation
\begin{equation}
    (P^2 - m^2) \Phi = 0.
\end{equation}
Here the Pauli-Lubanski operator is null, since the generators $\tensor{M}{^\mu^\nu}$\ map to the zero matrix, and there are no more constrictions to impose.  

Now consider the non--chiral spin one representation $\procarep$. A basis for the vector space is given by
\begin{equation}
    \left\{\ket{\psi_i} \right\} = \left\{ \ket{\tfrac{1}{2},\tfrac{1}{2}}, \ket{\tfrac{1}{2}, -\tfrac{1}{2}} , \ket{-\tfrac{1}{2},\tfrac{1}{2}} , \ket{-\tfrac{1}{2},-\tfrac{1}{2}}  \right\}.
\end{equation}
To analyze the operator space, we first note that every operator can be written in terms of the basis as some lineal combination:
\begin{equation}
    \mathcal{O} = \sum_{i,j} \tensor{\mathcal{O}}{_i_j} \ket{\psi_i} \bra{\psi_j}.
\end{equation}
The properties of the operators $\mathcal{O}$ are inherited from the properties of the states $\ket{\psi_i}$. As has been previously shown by one of the authors, the operator space will decompose as the square of the vector space \citep{2013Gómez-Ávila}:
\begin{equation}
    \procarep^{\otimes 2} \simeq \lrep{0,0} \oplus \lrep{1,0}_\chi \oplus \lrep{1,1}.
\end{equation}
Equivalently, 
\begin{equation}
    4 \otimes 4 \simeq 1 \oplus( 3 \oplus \bar{3}) \oplus 9.
\end{equation}
This corresponds to a scalar, an antisymmetric tensor, and a traceless symmetric tensor: 
\begin{equation}
    \{\id, \tensor{M}{_\mu_\nu}, \tensor{T}{_\mu_\nu}   \}.
\end{equation}
Now there are two possibilities for the kinetic term: we can use the scalar $\lrep{0,0}$ or the symmetric traceless tensor $\tensor{T}{_\mu_\nu}$ transforming as $\lrep{1,1}$. (The antisymmetric tensor cannot provided a kinetic terms, since it vanishes on contraction with the momenta). In the most general case, we would write a  wave equation
\begin{equation}
    (\alpha P^2 + \beta \tensor{T}{_\mu_\nu} P^\mu P^\nu  - m^2) \Psi = 0
\end{equation}
with unknown coefficients $\alpha, \beta$. Now, the requisite that this propagates three degrees of freedom, as appropiate for a spin $j$ field, and the values of the mass squared $m^2$ , will fix the values of $\alpha, \beta$. Equivalently, this amounts to imposing the eigenvalue equations for the squared momentum and Pauli-Lubanski operators. In any case, we end up with Proca's equation for a spin $1$ vector field. 

Let us look now at the Dirac representation $\diracrep$. This is also a four dimensional representation, with basis
\begin{equation}
    \left\{\ket{\psi_i}\right\} = \left\{ \ket{\tfrac{1}{2},0}, \ket{-\tfrac{1}{2}, 0} , \ket{0, \tfrac{1}{2}} , \ket{0,-\tfrac{1}{2}}  \right\}.
\end{equation}

The operator space for this representation will have the following decomposition: 
\begin{equation}
    \left[ \left( \frac{1}{2}, 0 \right) \oplus \left( 0, \frac{1}{2} \right) \right]^2 
\simeq (0,0)_2 \oplus (1,0) \oplus (0,1) \oplus \left( \frac{1}{2}, \frac{1}{2} \right)_2,
\end{equation}
which we can also write as
\begin{equation}
    ( 2 \oplus \bar{2}) \otimes ( 2 \oplus \bar{2}) \simeq 1_2 \oplus( 3 \oplus \bar{3}) \oplus 4_2.
\end{equation}
This corresponds to a covariant basis with two scalars, an antisymmetric tensor, and two vectors:
\begin{equation}
\left\{ 1,\, \chi,\, M_{\mu\nu},\, S_{\mu},\, \chi S_{\mu} \right\}.
\end{equation}
(This vector is most commonly denoted by $\gamma_\mu$; here we avoid this to keep a consistent notation in all cases.) It bears noting that, because this is a representation of $\Oo(1,3)$ rather than the proper Lorentz group $\SO^+(1,3)$, there is an intrinsic notion of parity, which is the operator that exchanges the chiral $\lrep{a,b}$ and $\lrep{b,a}$ parts. In our basis this is simply the matrix
\begin{equation}
   \Pi = \begin{pmatrix}  0 & \id_{2\times2} \\ \id_{2\times2} & 0\\
    \end{pmatrix}.
\end{equation}

Starting from the rest--frame parity projection
\begin{equation}
    \frac{1}{2} (1 \pm \Pi)\, \psi(0) = \psi(0),
\end{equation}
and performing a boost into an arbitrary frame, we obtain the covariant equation:
\begin{equation}
\left( S_{\mu} P^{\mu} \pm m \right) \psi(p) = 0.
\end{equation}
In other words, the operator $S_\mu$ can be interpreted as a covariant form of the parity operator, or rather, as the set of operators that parity transforms into under the action of the Lorentz group. This is what is meant by saying that the Dirac equation is simply the covariant projection over parity invariant subspaces of the $\diracrep$ representation  \citep{2013Gómez-Ávila}.

\section{Parity-invariant constructions for spin \texorpdfstring{$j > \tfrac{1}{2}$}{j>1/2}.}
\label{sec:joosweinberg}

In the single--spin chiral representations, also called Joos-Weinberg theories \citep{Joos:1962qq, Weinberg:1964cn, Weinberg:1964ev}, there are two possible kinetic operators. To see this, let us separate in two parts the calculation of the operator space decomposition. First, we have block diagonal operators coming from the products:
\begin{equation}
    \lrep{j,0}\otimes \lrep{j,0} \oplus \lrep{0,j}\otimes \lrep{0,j} \simeq \lrep{0,0}_2 \oplus \bigoplus_{i=1}^{2j} \lrep{i,0}_\chi.
\end{equation}
Only the scalar operators in $\lrep{0,0}_2$ can give us a kinetic term, because the series $\lrep{1,0}_\chi, \lrep{2,0}_\chi,\dots$ corresponds to a series of tensors with up to $2j$ pairs of antisymmetric indices:
\begin{equation}
    \bigoplus_{i=1}^{2j} \lrep{i,0}_\chi = \yng(1,1) \oplus \yng(2,2) \oplus \yng(3,3) \oplus\dots 
\end{equation}

This scalar kinetic term will produce a Klein-Gordon equation of motion for every spinor component. The drawback of this approach is that this is tipically double the number of degrees of freedom desired, since this includes both parities. On the other hand, we also have two anti block diagonal operators:
\begin{equation}
    2(\lrep{j,0}\otimes \lrep{0, j}) \simeq \lrep{j,j}_2.
\end{equation}
This correspond to another possible kinetic term, formed from a symmetric traceless tensor with $2j$ indices. Therefore, this is of order $2j$ in the momenta; for $j>1$, it corresponds to a higher--derivative theory. This kinetic terms is the one obtained by boosting the eigenvalue equation of the parity operator in the rest frame \citep{2013Gómez-Ávila}. 

There is another possibility, the double spin chiral representation where spin $j$ fields are induced from the $\left(j-\frac{1}{2},\frac{1}{2}\right)_\chi$
representation. We only need to consider $j\geq \tfrac{3}{2}$, since for $j=1$ this merely reproduces the vector representation. Block diagonal operators in the direct--sum basis come from 
\begin{equation}
    \begin{aligned}
        \left(j-\frac{1}{2}, \frac{1}{2}\right)^{\otimes 2} &= \bigoplus_{r=0}^{2j-1} [ (r,0)\oplus(r ,1) ],\\  
        \left(\frac{1}{2},j-\frac{1}{2}\right)^{\otimes 2} &= \bigoplus_{r=0}^{2j-1} [ (0,r)\oplus(1,r) ].\\
    \end{aligned}
\end{equation}
Since $j\geq \tfrac{3}{2}$, this will always include a pair of symmetric tensors $\lrep{1,1}$. On the other hand,  anti block diagonal operators come from the cross product 
\begin{equation}
    2 \left(j-\frac{1}{2}, \frac{1}{2}\right)\otimes  \left(\frac{1}{2},j-\frac{1}{2}\right) = 2 \bigoplus_{r,s=j-1}^{j} (r,s). 
\end{equation}
In this case, the lowest order kinetic term comes from the anti diagonal product $(j-1,j-1)$, which means that we have a first-order wave equation for $j=3/2$, and a second-order wave equation for $j=2$. Higher spin will result again in a higher--derivative theory.   

Finally, we have the possibility of using a $(j,j)$ completely symmetric tensor, which produces a kinetic term of momentum order $2j$, and therefore a higher derivative theory if $j\geq \frac{3}{2}$. In the following sections, we discuss in more detail this possibilities for $j=\frac{3}{2}$.

\section{Spin \texorpdfstring{$\frac{3}{2}$}{3/2} fields in the JW representation}
\label{sec:sschir}

The Joos-Weinberg formalism using the single spin chiral representations started as an attempt to directly construct transition amplitudes without having a wave equation or a Lagrangian\citep{Weinberg:1964cn, Weinberg:1964ev, Ahluwalia:1999ny}. This does not mean, however, that wave equations are impossible to define for the Jooos-Weinberg representations \citep{ahluwalia1992}, as we will exhibit below. 

For spin $\frac{3}{2}$, the single spin chiral representation operator space decomposes as:
\begin{equation}
    \begin{aligned}
        \left[\sschir\right]^{\otimes 2} =\ & \left(0,0\right)_2\oplus \left(\frac{3}{2},\frac{3}{2}\right)_2  \oplus(1,0) \oplus (0,1) \\ & \oplus(2,0)\oplus(0,2)\oplus(3,0)\oplus(0,3),
    \end{aligned}
\end{equation}
corresponding to the Young diagrams
\begin{equation}
  \id_2 \oplus \yng(3)_{\ 2}  \oplus \yng(1,1) \oplus \yng(2,2) \oplus \yng(3,3)\ . 
\end{equation}
Equivalently, 
\begin{equation}
    \begin{aligned}
        (4'\oplus \bar{4'})^2 =\ & 2\times 1 \oplus 2\times 16 \oplus 3 \oplus \bar{3} \oplus 5 \oplus \bar{5} \oplus 7 \oplus \bar{7}.
    \end{aligned}     
\end{equation}

We identify this as the identity $\id$ and chirality $\chi$ scalars, a
pair of third rank totally symmetrical tensors, and the operators
\begin{equation}\begin{split}
    \{ \tensor{M}{_\mu_\nu}, \tensor{C}{_\mu_\nu _\rho _\sigma},
    \tensor{D}{_\mu_\nu _\rho _\sigma _\alpha_\beta} \}.
  \end{split}\end{equation}
The $C$ and $D$ tensor are given by the Young-projected product of 
\begin{equation}\label{Weyl}
  \begin{split}
    C_{\mu\nu\alpha\beta} &= \mathcal{P}_{\tiny\yng(2,2)}\tensor{M}{_\mu_\nu} \tensor{M}{_\alpha_\beta},\\
    D_{\mu\nu \rho \sigma \alpha\beta} & = \mathcal{P}_{\tiny\yng(3,3)} \tensor{M}{_\mu_\nu} \tensor{M}{_\rho_\sigma} \tensor{M}{_\alpha_\beta}.\\
  \end{split}
\end{equation}
 For the symetrical tensor: 
\begin{equation}
\begin{aligned}
S_{\mu\nu\rho} &= \frac{1}{2}\Pi(-\eta_{0\mu}\eta_{0\nu}\eta_{0\rho} + \eta_{\mu\nu}\eta_{0\rho} + \eta_{\mu\rho}\eta_{0\mu} + \eta_{\rho\nu}\eta_{0\mu}) \\
&\quad + \frac{i}{9}\Pi\left[7{\eta_{\mu\nu}}M_{0\rho} + 7{\eta_{\mu\rho}}M_{0\nu} + 7{\eta_{\nu\rho}}M_{0\mu}\right] \\
&\quad - i\Pi\left(\eta_{0\mu}\eta_{0\nu}M_{0\rho} + \eta_{0\mu}\eta_{0\rho}M_{0\nu} + \eta_{0\nu}\eta_{0\rho}M_{0\mu}\right) \\
&\quad + \frac{2i}{9}\Pi(M_{0\mu}, M_{0\nu}, M_{0\rho}).
\end{aligned}
\end{equation}

With the covariant basis in hand, we can list all Lorentz covariant bilinear terms and use them to build kinetic terms and interactions. The Lagrangian density for a field transforming in this representation can contain only the following kinetic terms:
\begin{equation}
    \lagrangian_K =  \partial^\mu \overline{\Psi}(A + B \chi)\tensor{S}{_\mu_\nu_\rho} \partial^\nu \partial^\rho\Psi + (E + F\chi) \partial_\mu \overline{\Psi} \partial^\mu \Psi.
\end{equation}-

This will produce one of three outcomes: a second order, Klein--Gordon--like equation (with the disadvantage of fermion doubling, since both parities are slutions) a third order boosted--parity equation, or a mixture, which can be treated as in work by \citet*{Gristein2008a} and \citet*{Anselmi_FAKEONS}.

As for the double spin chiral representation, it has been mostly studied as a component of the Rarita-Schwinger representation \citep{Rarita:1941mf}. Its covariant space decomposes as
\begin{equation}
    \begin{aligned}
        \left[\dschir\right]^{\otimes 2} =\ & \left(0,0\right)_2 \oplus \left(\frac{1}{2}, \frac{1}{2} \right)_2 \oplus (1,1)_2 \oplus \left(\frac{3}{2}, \frac{3}{2} \right)_2 \\ &
        \oplus [(1,0)\oplus (0,1)]_2   \oplus (2,0) \oplus (0,2) \\ & \oplus \left(\frac{3}{2},\frac{1}{2}\right)_2 \oplus  \left(\frac{1}{2},\frac{3}{2}\right)_2 \oplus \left(2,1\right) \oplus  \left(1,2\right),
    \end{aligned}
\end{equation}
corresponding to the Young diagrams
\begin{equation}
    \begin{aligned}
          & \id_2 \oplus \yng(1)_{\ 2} \oplus \yng(2)_{\ 2} \oplus \yng(3)_{\ 2} \\ &  \oplus \yng(1,1)_2 \oplus \yng(2,2) \oplus \yng(2,1)_2 \oplus \yng(3,1) \ .     
    \end{aligned}
\end{equation}
Equivalently, 
\begin{equation}
    \begin{aligned}
        (6\oplus \bar{6})^2 =\ & 2\times 1 \oplus 2\times 4 \oplus 2\times 9 \oplus 2\times 16  \\
                              & \oplus 2\times(3 \oplus \bar{3}) \oplus 5 \oplus \bar{5} \oplus 2\times(8 \oplus \bar{8}) \oplus 15 \oplus \bar{15}.
    \end{aligned}     
\end{equation}

Focusing on the possible kinetic terms, and modulo the presence of the chirality operator $\chi$, we have four covariant possibilities:
\begin{equation}
    \{ \id, \beta^\mu,  \tensor{T}{^\mu^\nu}, \tensor{U}{^\rho^\mu^\nu} \},
\end{equation}
producing a first, second or third order wave equation. 

\section{Duffin-Kemmer-Petiau and the meson algebra}
\label{sec:meson algebra}
The  Duffin-Kemmer-Petiau \citep{1936Petiau, Duffin1938, 1939Kemmer},  construction originated in de~Broglie's \emph{neutrinic theory of light}, which treated the photon as a neutrino--antineutrino bound state. It is a linear wave equation analogous in structure to Dirac's formalism for fermions:
\begin{equation}
\left( i \beta^{\mu} \partial_{\mu} - m \right) \Psi = 0.
\end{equation}

The meson algebra, satisfied by these matrices, served as a generalization of Dirac's gamma matrices algebra \citep{Duffin1938}:
\begin{equation}
\beta^\mu \beta^\lambda \beta^\nu + \beta^\nu \beta^\lambda \beta^\mu = \beta^\mu \eta^{\lambda\nu} + \beta^\nu \eta^{\lambda\mu}.
\end{equation}
There are $\underline{5}$ and $\underline{10}$ dimensional representations of this meson algebra. \citet*{1974Hurley546} showed that these can be written as
\begin{equation}\label{eq:dkp3p1}
    \begin{aligned}
        \underline{5} &\simeq \lrep{0,0} \oplus \lrep{\frac{1}{2}, \frac{1}{2}} \\
        \underline{10} &\simeq \lrep{\frac{1}{2}, \frac{1}{2}} \oplus \lrep{1,0}_\chi\\
    \end{aligned}
\end{equation}
in terms of their Lorentz transformation rules. This DKP theory was used in the calculation of various nuclear processes involving vector and pseudoscalar mesons \citep{Clark1985, Klbermann1986, Kozack1989}. 

This is a puzzling construction from the point of view of our previous discussion, since it does not use a single representation of $\Oo(1,3)$, but rather a direct sum of chiral and nonchiral representations. Nevertheless, it provided an algebraically consistent construction, preserving the causality properties for spin one particles without subsidiary conditions. While these spin zero and one constructions can be shown in the free case to be equivalent to the better known Klein Gordon and Proca formalisms \citep{BouchefraBoudjedaa}, they are inequivalent when interactions are present \citep{Scadron1974}.

Looking to generalize the DKP construction to other spins, \citet*{1974Hurley546} studied all possible linear wave equations where the coefficients satisfied the condition
 \begin{equation}
     (\beta^\mu P_\mu)^3 = P^2 (\beta^\mu P_\mu), 
 \end{equation}
They concluded that such a linear equation existed only for the families of representations listed below (and their parity--conjugates): 
\begin{enumerate}
    \item $(n, 0) \oplus \left(n-\frac{1}{2}\frac{1}{2}\right)$,
    \item $(n, 0) \oplus \left(n+\frac{1}{2}, \frac{1}{2}\right)$,
    \item $\left(n+\frac{1}{2} \frac{1}{2}\right) \oplus (n,0) \oplus \left(n-\frac{1}{2}, \frac{1}{2}\right)$,
    \item $(1,0) \oplus \left(\frac{1}{2}\frac{1}{2}\right) \oplus (0,1)$.
\end{enumerate} 
 
Coming back to the double--spin chiral representation for $j=\frac{3}{2}$, where a vector operator was available to construct a linear wave equation, we note that this representation is not in the list. Since this representation has a left and a right component, in the direct sum basis it it possible to classify the operators in block diagonal and block antidiagonal parts. The block diagonal operators come from the left--left and right--right projectors corresponding to the Young diagrams
\begin{equation}
    \begin{aligned}
          & \id_2  \oplus \yng(2)_{\ 2}   \oplus \yng(1,1) \oplus \yng(2,2) \oplus \yng(3,1) \ .
    \end{aligned}
\end{equation}
while the block anti--diagonal operators are those in 
\begin{equation}
    \begin{aligned}
          & \yng(1)_{\ 2} \oplus \yng(3)_{\ 2} \oplus \yng(2,1)_2 \ .     
    \end{aligned}
\end{equation}
This is a $\mathcal{Z}_2$--gradation of the algebra; it implies  that for either of the vector operators, 
\begin{equation}\label{eq:grad}
    \beta^\mu \beta^\nu \beta^\rho  \sim  A \tensor{\eta}{^\mu ^\nu} \beta^\rho + B  \tensor{U}{^\rho^\mu^\nu} + C \tensor{G}{^\rho^\mu^\nu},
\end{equation}
 since the product of three block anti--diagonal matrices must itself be block anti--diagonal. 
 
By explicit calculation we see that for the $\lrep{1,\frac{1}{2}}_\chi$ representation,
\begin{equation}
    (\beta^\mu P_\mu) (\beta^\nu P_\nu) (\beta^\rho P_\rho) = A P^2 (\beta^\mu P_\mu) + B \tensor{U}{^\rho^\mu^\nu} P_\mu P_\nu P_\rho,
    \label{condition}
\end{equation}
with $B$ nonzero. This means that this linear equation is not quite analogous to Dirac's, and that the vector operator does not satisfy a meson algebra. We also point out that some of the representations in this list have not been fully explored in the literature as possibilities for the construction of relativistic wave equations.

\section{Covariant vector algebras from field theories in \texorpdfstring{$1+4$}{1+4}.}
\label{sec:xdredux}

A mathematical property that might shed light on the Duffin-Kemmer-Petiau construction is the embedding of the DKP representations in irreducible representations of a higher-dimensional Lorentz algebra. In fact, the $\SO(1,4)$ group has irreducible representations of dimension $\underline{5}$ and $\underline{10}$ which reduce precisely to the representations in \refeq{eq:dkp3p1}:
\begin{equation}
  \label{eq:detbranching5}
  \begin{aligned}
    \underline{5} & \hookrightarrow 1 \oplus 4\\
    \underline{10} & \hookrightarrow 3 \oplus \bar{3} \oplus 4.
  \end{aligned}
\end{equation}
If we we were to construct a wave equation for this extradimensional fields, we would start by calculating the covariant basis of operators. In the case of the $\underline{10}$ irrep, this basis is spanned by operators transforming as
\begin{equation}
  \label{eq:10covbasis}
  \underline{10}^2\simeq \underline{1}\oplus \underline{5}\oplus \underline{10} \oplus \underline{14} \oplus \underline{35} \oplus \underline{35}'. 
\end{equation}

This corresponds, in 4+1 language, to a scalar, a vector, a  symmetric tensor, and antisymmetric tensor, a mixed-symmetry tensor with the symmetries of the Christoffel connection, and a traceless tensor with the symmetries of the Weyl tensor, (all traceless), or in Young diagrams,
\begin{equation}
  \label{eq:10x10}
  \yng(1,1) \otimes \yng(1,1)  =
  \id  \oplus \yng(1) \oplus \yng(1,1)
  \oplus \yng(2) \oplus \yng(2,1)
  \oplus \yng(2,2).
\end{equation}

We can give a covariant description of the vector operator $\underline{5}$  in terms of the Lorentz algebra generators in
$4+1$ dimensions, $\tensor{M}{_\mu_\nu}$:
\begin{equation}
  S_\mu \equiv \frac{1}{2^3}\tensor{\eta}{_\mu_\nu}\tensor{\epsilon}{^\nu^\alpha ^\beta ^\rho ^\sigma} \{M_{\alpha \beta}, M_{\rho \sigma}\}.
\end{equation}
This vector obeys the five--dimensional version of the meson algebra. By looking at this five--dimensional field theory with four--dimensional glasses, we recover the DKP equation. 

This intriguing extradimensional connection leads us to ask if there are other cases where dimensional reduction can produce novel wave equations, particularly for spin $\frac{3}{2}$. We do not require that they give rise to linear wave equations. There turn out to be two different irreps of $\SO(1,4)$ which upon dimensional reduction produce spin $\frac{3}{2}$ as the highest in the representation:

\begin{equation}
  \label{eq:detbranchingegr}
  \begin{aligned}
    \underline{16} & \hookrightarrow 6 \oplus \bar{6} \oplus 2 \oplus  \bar{2}\\
    \underline{20} & \hookrightarrow 6 \oplus \bar{6} \oplus 4' \oplus  \bar{4'}.
  \end{aligned}
\end{equation}
The first one of these is the well-known Rarita-Schwinger representation, while the second one is a direct sum of the single and double spin chiral representations previously discussed. The Rarita-Schwinger equation is in this sense analogous to the DKP spin one formalism, a similitude already pointed out by \citet*{RKLoide_1997}.

\section{5-dimensional spin \texorpdfstring{$\frac{3}{2}$}{3/2} fields}
The $ \SO(1,4)$ $16$ Rarita--Schwinger representation has the following covariant basis decomposition:
\begin{equation}
    16\otimes 16 = 1 \oplus 5 \oplus 10_2 \oplus 14 \oplus 30 \oplus 35_2 \oplus 35' \oplus 81. 
\end{equation}

This corresponds to the Young diagrams
\begin{equation}
    \begin{aligned}
          & \id \oplus \yng(1) \oplus \yng(1,1)_{\ 2} \oplus \yng(2) \oplus \yng(3) \\ &  \oplus \yng(2,1) \oplus \yng(2,2) \oplus \yng(3,1) \ .     
    \end{aligned}
\end{equation}
The branching rules for the decomposition of these into $3+1$ Lorentz representation are
\begin{equation}
  \label{eq:detbranchingop16}
  \begin{aligned}
     1 & \hookrightarrow 1 \\
     5 & \hookrightarrow 1 \oplus 4 \\.
    10 & \hookrightarrow 3 \oplus \bar{3} \oplus 4 \\
    14 & \hookrightarrow 1 \oplus 4 \oplus 9 \\
    30 & \hookrightarrow 1 \oplus 4 \oplus 9 \oplus 16\\
    35 & \hookrightarrow 3 \oplus \bar{3} \oplus 4 \oplus 8 \oplus \bar{8} \oplus 9\\
    35' & \hookrightarrow 5 \oplus \bar{5} \oplus 8 \oplus  \bar{8} \oplus 9 \\
    81 & \hookrightarrow 3 \oplus \bar{3} \oplus 4 \oplus 8 \oplus \bar{8} \oplus 9 \oplus 15 \oplus \bar{15}  \oplus 16 \\
  \end{aligned}
\end{equation}
As we can see, there are multiple four--dimensional vectors here. However, most of them are coming from  higher dimensional
operators. This means that, if we were to build its kinetic terms with it in a Kaluza-Klein--type reduction only the extradimensional modes would obey a linear equation, while the zero--modes would have a second order kinetic term. (This is analogous to what happens in the five--dimensional DKP construction).
In purely four--dimensional language, we can choose from first and second order kinetic terms when building our wave equations. 

On the other hand, the $ \so(1,4)$ $20$ representation has the following covariant basis decomposition:
\begin{equation}
    20\otimes 20 = 1 \oplus 5 \oplus 10 \oplus 14 \oplus 30 \oplus 35 \oplus 35' \oplus 81 \oplus 84 \oplus 105,
\end{equation}
corresponding to the Young diagrams
\begin{equation}
    \begin{aligned}
          & \id \oplus \yng(1) \oplus \yng(1,1) \oplus \yng(2) \oplus \yng(3) \oplus \yng(2,1) \\ &   \oplus \yng(2,2) \oplus \yng(3,1) \oplus \yng(3,3) \oplus \yng(3,2) \ .     
    \end{aligned}
\end{equation}
 
The branching rules in this representation for the decomposition of these into $3+1$ Lorentz representation are the following:
\begin{equation}
  \label{eq:detbranchingop20}
  \begin{aligned}
     1 & \hookrightarrow 1 \\
     5 & \hookrightarrow 1 \oplus 4 \\.
    10 & \hookrightarrow 3 \oplus \bar{3} \oplus 4 \\
    14 & \hookrightarrow 1 \oplus 4 \oplus 9 \\
    30 & \hookrightarrow 1 \oplus 4 \oplus 9 \oplus 16\\
    35' & \hookrightarrow  3\oplus \bar{3} \oplus 4  \oplus  8\oplus \bar{8} \oplus 9\\
    35' & \hookrightarrow  5 \oplus \bar{5} \oplus 8 \oplus  \bar{8} \oplus 9\\
    81 & \hookrightarrow 3 \oplus \bar{3} \oplus 4 \oplus 8 \oplus \bar{8} \oplus 9 \oplus 15 \oplus \bar{15}  \oplus 16\\
    84 & \hookrightarrow 7 \oplus \bar{7}  \oplus 12 \oplus \bar{12} \oplus 15 \oplus \bar{15} \oplus 16\\
    105 & \hookrightarrow  5 \oplus \bar{5} \oplus 9 \oplus  8 \oplus \bar{8} \oplus 12 \oplus \bar{12} \oplus 15 \oplus \bar{15} \oplus 16 \\
  \end{aligned}
\end{equation}
There are multiple vectors providing linear kinetic terms, as well as second and third rank symmetric traceless tensors. As noted in previous work by one of the authors, the covariant parity eigenvalue equation will involve this third-rank symmetric tensor \citep{2013Gómez-Ávila}.

\section{Conclusions}
We have classified the Lorentz group representations giving rise to wave equations for spin $\frac{3}{2}$, subject to the  consistency conditions that no more that two spin sectors are obtained, and that the spin wanted be the highest present. We have found four possibilities:
\begin{enumerate}
    \item Single spin chiral representation $\sschir$, where we find a higher--derivative theory. 
    \item Double spin chiral representation $\dschir$ giving rise to a linear, quadratic or cubic wave equation.  
    \item The well-known Rarita-Schwinger representation $\rsr$, where we also find a linear, quadratic or cubic wave equation. 
    \item The novel $\egr$ representation, mixing the single and double spin chiral representations, and with first, second and third rank kinetic terms. 
\end{enumerate}

The study of the novel DKP--type theory in the $\egr$ representation is an interesting perspective of this work. To our knowledge, wave equations defined in this representation have not been previously studied in the literature. 

\bibliographystyle{unsrtnat}
\bibliography{waveeq}

\end{document}